\documentclass[twocolumn,showpacs,preprintnumbers,amsmath,amssymb,prb]{revtex4}

\usepackage{graphicx}% Include figure files
\usepackage{dcolumn}% Align table columns on decimal point
\usepackage{bm}% bold math
\usepackage{amsmath,amssymb}

\newcommand{\be}{\begin{equation}}
\newcommand{\ee}{\end{equation}}

\newcommand{\beq}{\begin{eqnarray}}
\newcommand{\eeq}{\end{eqnarray}}

\def\eq#1{(\ref{#1})}
\def\H1{\widehat{H}_1}

\newcommand{\pd}{\partial}

\begin{document}

\title{Crossover from diffusive to non-diffusive dynamics in the two-dimensional electron gas with Rashba spin-orbit coupling}

\author{M. Pletyukhov}
\affiliation{Institut f\"ur Theoretische Festk\"orperphysik, Universit\"at
Karlsruhe, D-76128 Karlsruhe, Germany and \\
Center for Functional Nanostructures, Universit\"at Karlsruhe, D-76128 Karlsruhe, Germany}

\begin{abstract}
We present the calculation of the density matrix response function of the two-dimensional electron gas with Rashba spin-orbit interaction, which is applicable in a wide range of parameters covering the diffusive and non-diffusive, the dirty and the clean limits. A description of the crossover between the different regimes is thus provided as well. On the basis of the derived microscopic expressions we study the propagating charge and spin-polarization modes in the clean, non-diffusive regime, which is accessible in the modern experiments.
\end{abstract}

\maketitle

\section{Introduction}

One of the important directions of study in the field of spintronics \cite{fabian} is a possibility to manipulate spin-polarized electrons by means of electric fields. In two-dimensional heterostructures or quantum wells, this can be achieved \cite{datta} due to Rashba spin-orbit (SO) interaction \cite{rash} which arises from the structure inversion asymmetry and couples linearly the electron's momentum to its spin. 

The standard approaches to the study of the charge and spin dynamics in impure two-dimensional electron gas (2DEG) with Rashba SO coupling is based on either Kubo formalism \cite{bur1} or  quantum kinetic equation \cite{shyt1} (QKE). In the diffusive limit $v_F q, \omega \ll \tau^{-1}$, where $\tau$ is an elastic mean free time, the usually made approximation is a small-$q$ expansion of the so-called diffusion propagator (or, equivalently, the angle averaging of the QKE). As a result, one ends up with the diffusion equation,\cite{bur1,shyt1} which along with the Dyakonov-Perel spin relaxation mechanism,\cite{dyak} accounts for the spin precession as well as for the dynamic spin-charge mixing. 

Such description applies at the large distances and time scales. The study of the charge- and spin-density evolution on short ($v_F q, \omega \gg \tau^{-1}$) or intermediate ($v_F q, \omega \sim \tau^{-1}$) scales requires either a more refine consideration of the density matrix response function or a solution of the QKE in the most general case. The recent progress in experimental observation of the non-diffusive spin dynamics \cite{weber} in the transient spin grating experiments  \cite{cameron,weber1} makes desirable the further elaboration of theoretical tools applicable beyond the diffusive limit. 

While comparing the above mentioned approaches, we would like to remark that the QKE is the differential equation for the local, non-stationary distribution function\cite{rammer} (more precisely, for the Keldysh component of a Green's function), which in the presence of the SO coupling has been recently considered in the spatially nonuniform situations in Refs.~\onlinecite{shyt2} and \onlinecite{schwab}. In the quasi-equilibrium regime the QKE approach can be  complemented by the alternative, but equivalent, consideration of the charge and spin densities in the Fourier representation. This seems to be especially useful in the context of the boundary  problem, which appears to be a subtle issue of the QKE approach in the clean, non-diffusive regime\cite{schwab}. 

In this paper we present the compact analytical expressions for the density matrix response function of the impure 2DEG with Rashba SO coupling. They cover quite a wide range of parameters, extending from the scale of  the diffusive transport to the arbitrary spatial and temporal modulations of external fields as well as to the  arbitrary values of the cleanness parameter $\alpha_R \tau k_F$, $\alpha_R$ being the Rashba constant. On this ground the description of crossover from the diffusive to non-diffusive (ballistic) regimes becomes quite efficient. In particular, we apply the obtained expressions to the study of propagating charge and spin-polarization modes, and discuss the $q$-dependence of their dispersions and lifetimes.

\section{Charge and spin density response functions}

The Hamiltonian of the 2DEG with Rashba SO coupling \cite{rash} in the absence of impurities reads
\be
H ({\bf k})= \frac{{\bf k}^2}{2 m} +\alpha_R (\sigma^x k_y - \sigma^y k_x),
\label{rham}
\ee
where $m$ is an effective mass, and $\hbar =1$ in the units used. The spectrum of \eq{rham} is given by $\varepsilon_{{\bf k}}^{\mu} = \frac{|{\bf k}|^2}{2 m} + \mu \alpha_R |{\bf k}|$, where $\mu=\pm$ is a subband index. The projectors onto the corresponding eigenstates are found to be
\be
P_{\mu} ({\bf k}) = \frac12 \left( \begin{array}{cc} 1 & i \mu e^{-i \phi_{{\bf k}}}\\ -i \mu e^{i \phi_{{\bf k}}} & 1 \end{array}\right), \quad \tan \phi_{{\bf k}} = \frac{k_y}{k_x}.
\label{proj}
\ee
Using \eq{proj}, one can perform a spectral decomposition of the Hamiltonian 
\eq{rham}:
\be
H ({\bf k}) = \sum_{\mu = \pm} P_{\mu} ({\bf k}) \varepsilon_{{\bf k}}^{\mu}.
\label{specdec}
\ee
A weak disorder is modelled by random isotropic point-like scatterers with Gaussian configuration distribution, and its strength is parametrically given by $( m \tau)^{-1}$.

For the future reference we introduce the following notations for the Fermi momentum $k_F = \sqrt{2 m E_F + k_R^2}$, the Fermi velocity $v_F=\frac{k_F}{m}$, and the Fermi energy $E_F$; as well as for  the Rashba momentum splitting $k_R=m \alpha_R$ and the density of states per spin component
$\nu = \frac{m}{2 \pi}$. When appropriate, we will also use the dimensionless units $y = \frac{k_R}{k_F}$, $z=\frac{q}{2 k_F}$, $\tilde{w} = w+i \Gamma= \frac{m}{2 k_F^2}(\omega+ \frac{i}{\tau}) \equiv \frac{m}{2 k_F^2} \tilde{\omega}$.

Like in Ref.~\onlinecite{shnir1}, we can define the following regimes depending on the relative strength of SO coupling and disorder potential: dirty $y \ll \Gamma$, clean $y^2 \leq \Gamma \ll y$, and super-clean $\Gamma \leq y^2$. The last regime is hardly accessible experimentally by now, and it will not be discussed in this paper. However, it is important to note that the clean regime as it is introduced here implies the limitations on $\Gamma$ from below as well as from above. The necessity to distinguish between the clean and the super-clean regimes will be discussed in the following.

A linear response of a system $\rho$ to an external perturbation $V$ is determined by the density response function $\chi = \chi ({\bf q}, \omega)$. In case of spin-resolved spectrum, $\chi= (\chi^{\alpha \beta})$ is a $4\times 4$-matrix ($\alpha,\beta =0,x,y,z$; the index ``0'' corresponds to the charge component), which includes the polarization operator ($\alpha, \beta =0$), the magnetization ($\alpha, \beta =x,y,z$), and the spin-charge mixing ($\alpha =0$, $\beta=x,y,z$; or vice versa). Quite generally, $\chi$ is evaluated by means of the disorder-averaged Green's function technique \cite{mah}. In the self-consistent Born approximation one finds at zero temperature \cite{bur1} $\chi = \chi^I +\chi^{II}$, where 
\beq
\chi^{I} ({\bf q},\omega) = 2 i \nu \tau \int_{-\omega/2}^{+\omega/2} d \varepsilon I ({\bf q}, \omega; \varepsilon) [1-I ({\bf q}, \omega; \varepsilon)]^{-1},
\label{deps} \\
I^{\alpha \beta} ({\bf q}, \omega; \varepsilon) = \frac{1}{4 \pi \nu \tau} \sum_{\mu,\mu'=\pm} \int \frac{d^2 k}{(2 \pi)^2} \nonumber \\
\times  \frac{{\cal F}^{\alpha \beta}_{\mu \mu'} ({\bf k},{\bf k}+{\bf q})}{(E_F +\varepsilon -\frac{\tilde{\omega}}{2} - \varepsilon_{{\bf k}}^{\mu}) (E_F +\varepsilon + \frac{\tilde{\omega}}{2}- \varepsilon_{{\bf k}+{\bf q}}^{\mu'})},
 \label{momint} 
\eeq
and ${\cal F}^{\alpha \beta}_{\mu \mu'} ({\bf k},{\bf k}+{\bf q})= {\rm Tr} [\sigma^{\beta} P_{\mu} ({\bf k}) \sigma^{\alpha}  P_{\mu'} ({\bf k}+{\bf q})]$, $\sigma^0 \equiv 1$,  $\tilde{\omega} = \omega + \frac{i}{\tau}$, as well as $\chi^{II, \alpha \beta} = 2 \nu \delta^{\alpha \beta}$. A careful analysis shows 
that
\be
\chi^{I} ({\bf q}, \omega) \approx 2 i \nu \omega \tau I ({\bf q}, \omega) [1-I ({\bf q}, \omega)]^{-1},
\label{diffapp}
\ee
where $I ({\bf q}, \omega) \equiv I ({\bf q}, \omega; \varepsilon=0)$, remains a good approximation in the dirty ($y \ll \Gamma$), intermediate ($y \sim \Gamma$), and clean ($y^2 \leq \Gamma \ll y$) regimes. It breaks down only in the super-clean regime, $y^2 \geq \Gamma$, and in this case the integration over $\varepsilon$ in Eq.~\eq{deps} has to be performed exactly. On the other hand, in the super-clean limit the disorder potential can be neglected at all ($\Gamma \to 0$). Then, following the approach of Ref.~\onlinecite{plet1}, one can find exact analytic expressions for $\chi^{\alpha \beta}$ at arbitrary $q$ and $\omega$ in this regime.

The vertex corrections to $\chi$ are accounted in \eq{deps} and \eq{diffapp} in the diffusion propagator, or diffuson, ${\cal D} ({\bf q}, \omega) = \tau [1 - I({\bf q}, \omega) ]^{-1}$. In the diffusive regime $v_F q, \omega \ll \tau^{-1}$, Eq.~\eq{momint} can be expanded in a series of small $q$ and $\omega$, which makes straightforward the evaluation of $I ({\bf q}, \omega)$ and ${\cal D} ({\bf q}, \omega)$ \cite{bur1}. The Fourier transform of ${\cal D}^{-1}$ gives rise to the diffusion equation \cite{bur1,shyt1} describing the evolution of charge and spin densities at large distances and time scales. In turn, a consideration of the short-range dynamics requires a more careful evaluation of the momentum integral in \eq{momint}, especially of its angular part. 

Before proceeding further in evaluation of $\chi$, let us establish the relation between the discussed diagrammatic approach and the approach based on the quantum kinetic equation\cite{shyt1}. Both of them can be deduced from the Dyson equation\cite{rammer}. We refer to the Appendix \ref{dyssect} where the corresponding alternative derivation of the response functions \eq{deps},\eq{momint} is presented. Note that the quasiclassical conditions  $q \ll k_F$ and $\omega \ll E_F$ have not been applied yet for the evaluation of $\chi$. This will be done in the next section.

In turn, the derivation of the QKE from the Dyson equation is based on the procedure which is specifically designed for the quasiclassical limit\cite{rammer}. The QKE represents the differential equation for the nonequilibrium distribution function. In the quasi-equilibrium (linear response) regime the QKE approach becomes fully equivalent to the quasiclassical linear response theory, since both are derived from the Dyson equation under the same assumptions. 

The solution of the QKE for the system in question also allows us to access the charge and spin dynamics in the clean and dirty, diffusive and non-diffusive cases. In the spatially nonuniform and non-diffusive situations the QKE in presence of the SO coupling has been previously considered in Refs.~\onlinecite{shyt2} and \onlinecite{schwab}. However, it is important to note that the time and space representation of the local distribution function requires a proper choice of the boundary conditions, which in presence of the strong SO coupling and weak disorder becomes a nontrivial problem. For example, in Ref.~\onlinecite{schwab} the boundary conditions imposed on the distribution function in the clean, non-diffusive case have been motivated by the phenomenological arguments borrowed from the scattering matrix theory, on the analogy with Ref.~\onlinecite{sauls}. We also refer to Ref.~\onlinecite{galits}, where similar issues have been analyzed in the diffusive regime.

The problem of the boundary conditions can be postponed (and, hopefully, circumvented), if the response functions are evaluated in the frequency and momentum representation. For example, applying the inverse Fourier transformation to a finite system of the size  $L \gg k_F^{-1}$, we would obtain the spatial dependence of the observable quantities $\rho^{\alpha} ({\bf r},t)$. The boundary conditions are thus effectively accounted in terms of $V^{\alpha} ({\bf r},t)$.

\section{Quasiclassical approximation}

The quasiclassical approximation of $\chi$ and $I$ [Eqs. \eq{deps} and \eq{momint}] is achieved due to  the substitution of the integral over $d |{\bf k}|$ in Eq.~\eq{momint} by a residue value. To this end, one can close a contour of integration in either upper or lower half-planes. However, a better approximation would be to take the mean value of the both results, because it supports the symmetry $I^{\dagger} ({\bf q}, \omega) = I (-{\bf q}, -\omega)$ of the exact expression \eq{momint} (at $\varepsilon =0$). Thus, we obtain the sum of the residia
\be
I^{\alpha \beta} \approx \sum_{\stackrel{\mu,\mu'}{\lambda = \pm}} \frac{i k \lambda^{\frac{1+s}{2}}}{8 \pi \tau  (k + \mu k_R)} \int_0^{2 \pi} \! \! \!  \! \! d \phi_{{\bf k}} \frac{{\cal F}^{\alpha \beta}_{\mu \mu'} ({\bf k}, {\bf k} +{\bf q})}{\lambda \tilde{\omega}+ \varepsilon^{\mu}_{{\bf k}}-\varepsilon^{\mu'}_{{\bf k}+{\bf q}}} \bigg|_{k=k^*_{\lambda,\mu}}
\label{angint}
\ee
calculated in the poles $k^*_{\lambda,\mu} = \sqrt{k_F^2 - \lambda m^* \tilde{\omega}} -\mu k_R$. The new index $\lambda$ labels the two different closures of the integration contour. Note that we also assume $k_R \ll k_F$, which is a physically justified condition.

In the basis ${\bf q} || {\bf e}_x$, the integration angle $\phi$ is measured from the direction of ${\bf q}$, and all non-vanishing components (see the Appendix \ref{integrat}) are distributed into the two blocks: $(I^{00},I^{yy}, I^{0y} = I^{y0})$ form the {\it even} block ($\alpha, \beta =0,y$), while $(I^{xx},I^{zz}, I^{xz} = -I^{zx})$ form the {\it odd} block ($\alpha, \beta =x,z$).  The parameter $s$ reflects the symmetry of ${\cal F}^{\alpha \beta}_{\mu \mu'} ({\bf k}, {\bf k} +{\bf q}) = s {\cal F}^{\alpha \beta}_{\mu', \mu} (-{\bf k}- {\bf q}, -{\bf k})$ under the three successive operations: ${\bf k} \to {\bf k} -{\bf q}$, ${\bf k} \to -{\bf k}$, and $\mu \leftrightarrow \mu'$. It appears to be that $s$ equals $1$ for all components, except for $s=-1$ for $I^{0y} = I^{y0}$.

The central point of our calculation is an exact integration over the angle variable in Eq.~\eq{angint}. The technical details of this evaluation are described in the Appendix \ref{integrat}. Here we present the final  result for the non-zero components of $I^{\alpha \beta}$:
\beq
I^{00} &=& I_0^{(e)}, \quad I^{0y} = I^{y0}= I_1^{(e)}, 
\quad I^{yy} = I_2^{(e)}, 
\label{ide} \\
I^{zz} &=& I_0^{(o)}, \quad I^{zx} = - I^{xz} = -i I_1^{(o)}, \quad I^{xx} = I_2^{(o)},
\label{ido}
\eeq
where 
\beq
I_n^{(e)} = \sum_{\mu=\pm} {\textstyle \frac{i \Gamma (-\mu)^n }{2 (1-\tilde{w}^2)}   \left( \frac{\zeta^{1-n}}{z_{\mu}} \sqrt{\frac{1-z_{\mu}^2}{1-\zeta^2}} - \frac{z^{1-n}}{\zeta_{\mu}} \sqrt{\frac{1 -\zeta_{\mu}^2}{1-z^2}}\right)}, 
\label{ice} \\
I_n^{(o)} = \sum_{\mu=\pm} {\textstyle \frac{i \Gamma (-\mu)^n}{2 (1-\tilde{w}^2)} \left( \frac{\zeta_{\mu}^{1-n}}{z} \sqrt{\frac{1-z^2}{1-\zeta_{\mu}^2}} - \frac{z_{\mu}^{1-n}}{\zeta} \sqrt{\frac{1-\zeta^2}{1-z_{\mu}^2}} \right)}.
\label{ico}
\eeq
In these expressions $z_{\mu} = \frac{z}{1 -\mu y}$, $\zeta=\frac{z}{\tilde{w}}$, $\zeta_{\mu}=\frac{z}{\tilde{w}-\mu y}$, and $n=0,1,2$. Note that $I_n^{(e)}$ and  $I_n^{(o)}$ are mapped onto each other, $I_n^{(e)} \leftrightarrow I_n^{(o)}$, when $\zeta \leftrightarrow \zeta_{\mu}$ and $z \leftrightarrow z_{\mu}$ are exchanged. Moreover, $I_n^{(e,o)} \to - I_n^{(e,o)}$ under the formal replacements $\zeta \leftrightarrow z$ and $z_{\mu} \leftrightarrow \zeta_{\mu}$.

The expressions \eq{ide}-\eq{ico} represent the main result of this paper. Their applicability range is determined by the conditions $y,\Gamma,z,w \ll 1$ and $y^2 \ll \Gamma$ which follow from the assumptions made in the course of the  calculation. As for the rest, they are equally well applicable in the diffusive ($z,w \ll \Gamma$) and non-diffusive ($z,w \gg \Gamma$) regimes, in the dirty ($y \ll \Gamma$) and clean ($y \gg \Gamma$) limits.

It is useful to consider different limiting cases of Eqs.~\eq{ide}-\eq{ico}, in order to prove their consistency with the well-established results. 

In the limit $y \to 0$, the off-diagonal terms vanish, $I^{0y}= I^{zx}=0$, while the diagonal terms $I^{\alpha \alpha}$ yield the  familiar expression  in the absence of SO coupling,
\be
I^{\alpha \alpha}_{y=0} = \frac{i \Gamma}{\sqrt{\tilde{w}^2 -z^2} \sqrt{1-z^2}}
\approx  \frac{i \Gamma}{\sqrt{\tilde{w}^2 -z^2}}.
\ee

In the homogeneous limit $z \to 0$, $I^{00} = \frac{i \Gamma}{\tilde{w}}$, $I^{zz} = \frac{i \Gamma \tilde{w}}{\tilde{w}^2 -y^2}$, and  $I^{xx}=I^{yy} = \frac12 (I^{00}+I^{zz})$. Setting $w=0$ in these expressions, we recover the out-of-plane $ \frac{1}{\tau_z} = \frac{1}{\tau} (1 -I^{zz}) = \frac{(y/\Gamma)^2}{\tau [1+(y/\Gamma)^2]}$ and the in-plane $\frac{1}{\tau_{\perp}} = \frac{1}{\tau} (1 -I^{xx}) = \frac{1}{\tau} (1 -I^{yy}) \equiv \frac{1}{2 \tau_z}$ spin relaxation rates of the Dyakonov-Perel mechanism \cite{dyak}. In turn, the  small-$z$ behavior of the off-diagonal components reads
\be
I^{0y} \approx -\frac{i y^3 \Gamma z}{2 \tilde{w}^2 (\tilde{w}^2 - y^2)}, \quad I^{zx} \approx - \frac{y \Gamma \tilde{w} z}{(\tilde{w}^2 - y^2)^2}.
\ee
At $w=0$, we recover the coefficients $\lim_{z \to 0} [i I^{0y}/( 2 k_F z)]$ and $\lim_{z \to 0} [ i I^{zx}/(2 k_F z)]$ of the diffusion equation describing the spin-charge mixing and the spin precession, respectively, which were previously derived in Refs.~\onlinecite{bur1,shyt1}. 

We note that the exact integration over the angle variable yields the expressions \eq{ice} and \eq{ico}, which contain the square roots of the complex-valued expressions with the branch-cuts along the lines $w =z$ and $w = |z\pm y|$. In the next section we will show that these branch-cuts determine dispersions of propagating modes in the clean, non-diffusive limit. Meanwhile, the perturbative expansion in $z$, which is only validated in the diffusive limit, destroys the square-root structure of the expressions  \eq{ice} and \eq{ico}, thus eliminating the branch-cuts from them. Therefore, one can hardly expect an occurrence of dispersive propagating modes in the diffusive regime.

\section{Structure factors and propagating modes}
\label{propag}

Diffusive modes and their lifetimes \cite{bur1} still remain in focus of an active study. They are naturally reproduced within the present approach in the dirty limit $y \ll \Gamma$ at $z,w \ll \Gamma$. However, there exists another interesting question about a probable occurrence of propagating modes in the clean regime $\Gamma \ll y$. For example, in Ref. \onlinecite{berv1} it was conjectured on the basis of quite general physical arguments that ``in the limit of strong spin-orbit coupling, there is a regime where a propagating, coupled spin-charge mode is possible.'' Now, the microscopically derived expressions \eq{ide}-\eq{ico} allow us  to tackle this issue as well as the issue of propagation of coupled $s^x$--$s^z$-polarization modes determined by the odd block of $\chi$. 

A convenient tool for studying the dispersions and lifetimes of either propagating or diffusive modes in all regimes is provided by the structure factor $S ({\bf q},\omega) = {\rm Tr}  \,\, {\rm Im} \chi ({\bf q},\omega)$, which is a directly measurable function \cite{mah}. In case of  Rashba SO coupling we can define the two block-resolved structure factors, $S^{(e)} = {\rm Tr}_e  {\rm Im} \chi$ and $S^{(o)} = {\rm Tr}_o {\rm Im} \chi$, such that $S = S^{(e)}+ S^{(o)}$. The peaks' positions of $S^{(e,o)}$ in the plane $(q, \omega)$ correspond to the dispersions of propagating (at $\omega=0$ -- diffusive) modes, while the peaks' widths determine the inverse lifetimes of the corresponding modes.

\begin{figure}[t]
%\begin{center}
\includegraphics[width=7.6cm,angle=0]{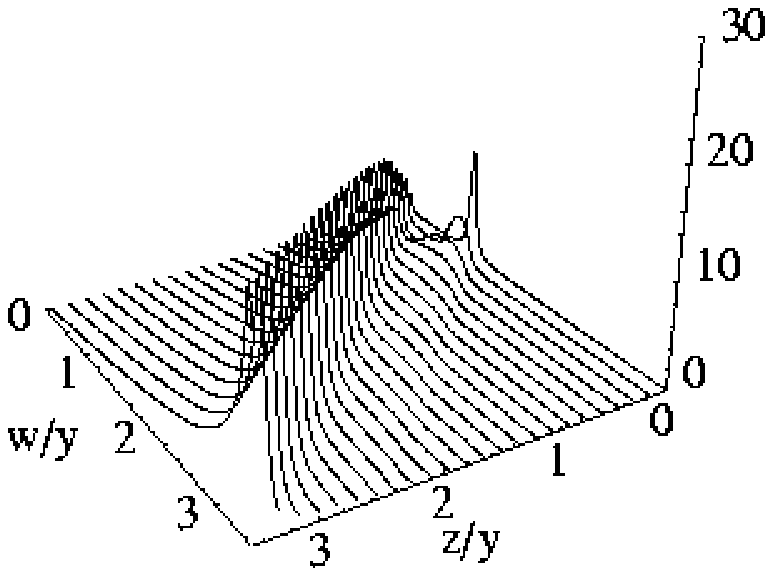} 
\includegraphics[width=7.6cm,angle=0]{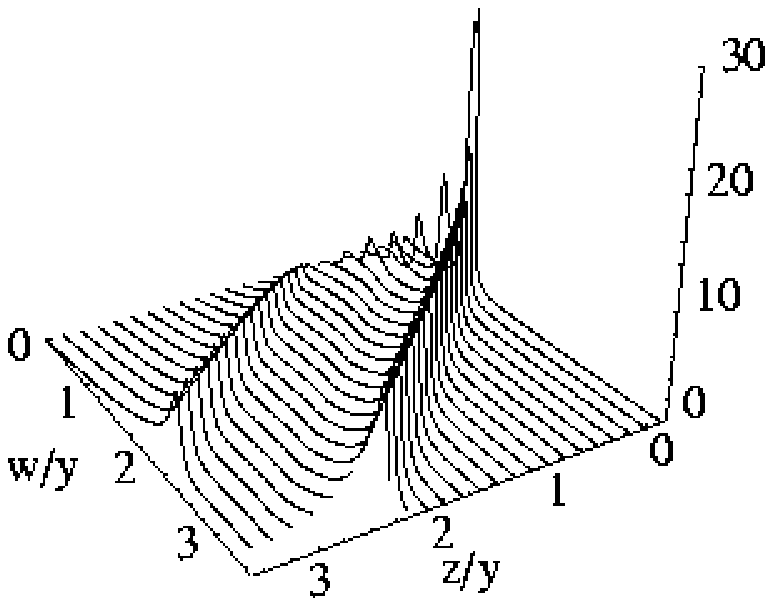}
\caption{The structure factors $S^{(e)} (z,w)/\nu$ (upper panel) and $S^{(o)} (z,w)/\nu$ (lower panel) at $y=10^{-3}$ and $\Gamma =0.5 \times 10^{-4}$.}
\label{struc}
%\end{center}
\end{figure}

In the upper and lower panels of the Fig.~\ref{struc} we plot $S^{(e)} (z,w)/\nu$ and $S^{(o)} (z,w)/\nu$, respectively. The SO and the disorder parameters are chosen to be $y=10^{-3}$ and $\Gamma = 0.5 \times 10^{-4}$. By the order of magnitude these values correspond to those of the high-mobility sample from Ref.~\onlinecite{weber} which was characterized by $\mu \approx 1.5 \times 10^5$ cm$^2$/V-s. Note that $y \gg \Gamma \gg y^2 = 10^{-6}$, i.e. the chosen parameters belong to the clean regime, but they are still far away from the super-clean regime. The values of the grating wave vector\cite{weber} $q = 0.44 -5.3 \times 10^4$ cm$^{-1}$ correspond to $z = 10^{-3} - 10^{-2}$, which means that the non-diffusive regime $z>\Gamma$ is accessible. However, a direct comparison with the experiment is obstructed by an admixture of the Dresselhaus contribution to the SO coupling, which is not taken into account in our consideration. 

In the Fig.~\ref{struc} one  can clearly see the dispersions of the four propagating modes: $w^{(e)}_+ \approx z$ and $w^{(e)}_- \approx y$ in the upper panel, and $w^{(o)}_+ \approx z+y$ and $w^{(o)}_- \approx |z-y|$ in the lower panel. We note that the fact of existence of the gapless mode $w^{(e)}_+ \approx z$ has been already pointed out in Ref.~\onlinecite{raimondi}  as the consequence of the charge conservation law. 

\begin{figure}[t]
%\begin{center}
\includegraphics[width=7.6cm,angle=0]{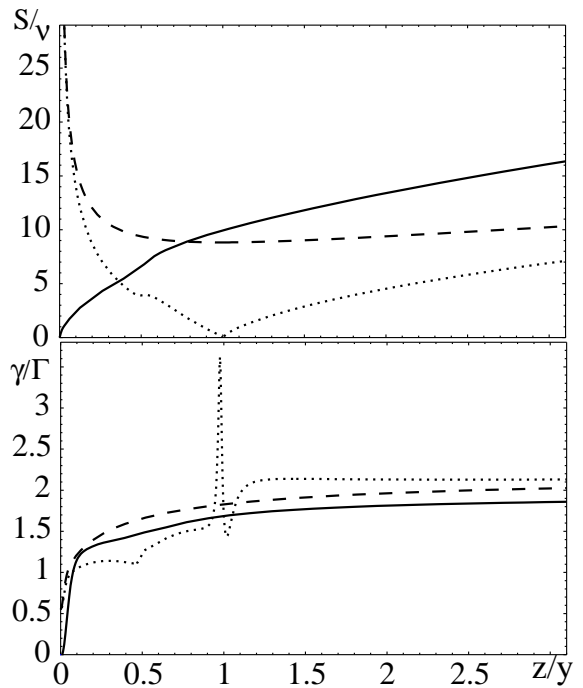}
\caption{{\it Upper panel:} the peaks' heights (weights) $S^{(o)} (z, w^{(o)}_{\pm} (z))$ (dashed  and dotted) and  $S^{(e)} (z, w^{(e)}_+ (z))$ (solid) normalized by $\nu$. {\it Lower panel:} the widths $\gamma^{(o)}_{\pm} (z)$ and $\gamma^{(e)}_+ (z)$ of the corresponding modes normalized by $\Gamma$.}
\label{combine}
%\end{center}
\end{figure}

In order to understand physical meaning of the other modes, let us  consider $S^{(e,o)} (z,w)$ at small and large $z$. 

In the homogeneous limit $z \to 0$, $w_{-}^{(e)} \approx w_{\pm}^{(o)} \approx y$, and we can approximate the structure factors by the Lorentzians,
\be
S^{(e,o)} (0, w) \approx \frac{2 \nu y}{\Gamma} \frac{\alpha^{(e,o)}_{-,\pm} (0) \gamma^{(e,o)}_{-,\pm} (0)}{(w-y)^2 + [\gamma^{(e,o)}_{-,\pm} (0)]^2},
\ee
with the weights $\alpha^{(e)}_- (0) =  \frac14 \Gamma$ and $\alpha^{(o)}_{\pm} (0) =  \frac{4}{\sqrt{31}} \Gamma$, and the widths $\gamma^{(e)}_- (0) =  \frac34 \Gamma$ and $\gamma^{(o)}_{\pm} (0)  =  \frac{3}{\sqrt{31}} \Gamma$.

With increasing $z$ the mode $w_{-}^{(e)}$ broadens very fast, simultaneously losing its weight. At the value $z \approx 0.5 y$ it is no longer resolved, but after this point its weight is augmented again. Thus, the saddle near the point $(z,w) \approx  (0.5 y, y)$ is formed in the landscape of $S^{(e)}$. At $z \approx y$ both $w_{-}^{(e)}$ and $w_{+}^{(e)}$ merge together, and for large $z \gg y$ the mode $w_{+}^{(e)}$ becomes degenerate, in the sense that the dispersions of propagating charge and $s^y$-polarization modes overlap. It also means that $S^{(e)} \equiv {\rm Im} (\chi^{00} + \chi^{yy}) \approx 2 {\rm Im} \chi^{00} \approx 2 S_0 (z,w)$, where  $S_0 (z,w) = - \nu {\rm Im}\sqrt{\frac{2 z }{w-z+i \Gamma}}$ is a large-$z$ structure factor in the absence of SO coupling. 

We note that even in the clean limit the value of $I^{0y}$ remains negligibly small ($|I^{0y} |\ll |I^{00}|, |I^{yy}|$) at any $z$ and $w$. By this reason, the spin-charge mixing can hardly be detected in the study of the propagating modes.

The modes $w^{(o)}_+$ and $w^{(o)}_-$ behave more peculiarly at finite $z$, especially the latter one. Both of them are coupled $s^x$--$s^z$-polarization modes originating in the point $(z,w) \approx (0,y)$. However, there is a qualitative difference between them: the mode $w^{(o)}_+$ grows monotonously with $z$, while the mode $w^{(o)}_-$ initially goes down towards $(z,w) \approx (y,0)$, where it turns up, and after that it parallels $w^{(o)}_+$. The asymptotic behavior of $S^{(o)}$ at large $z \sim w \gg y$ is $S^{(o)} (z,w) \approx S_0 (z,w-y) + S_0 (z,w+ y)$, i.e.  $w^{(o)}_{\pm}$ remain well-resolved and split-off of $w^{(e)}_+$. Note that the SO parameter $y$ determines only the amount of this splitting, but not the shape of the individual peaks.

The most interesting aspect seen in the lower panel of the Fig.~\ref{struc} is the form of the structure factor $S^{(o)}$ in the vicinity of $(z,w) \approx (y,0)$: it locally vanishes, which corresponds to the local suppression of the $w^{(o)}_-$ mode. This is also clearly manifested in the upper panel of the Fig.~\ref{combine}, where the absolute values (weights) of the structure factors  $S^{(e)} (z, z)$ and $S^{(o)} (z,|z\pm y|)$ are plotted. Using the Lorentzian approximation, we estimate the widths $\gamma^{(e,o)}_{+,\pm} (z)$ of the corresponding peaks, and present them in the lower panel of the Fig.~\ref{combine}. One can notice that near $z \approx y$, $\gamma_-^{(o)} (z)$ (dotted curve) shows up specific features which are, however, somewhat artificially created: since the corresponding peak locally vanishes, the Lorentzian approximation also becomes locally inadequate.

\begin{figure}[t]
%\begin{center}
\includegraphics[width=7.6cm,angle=0]{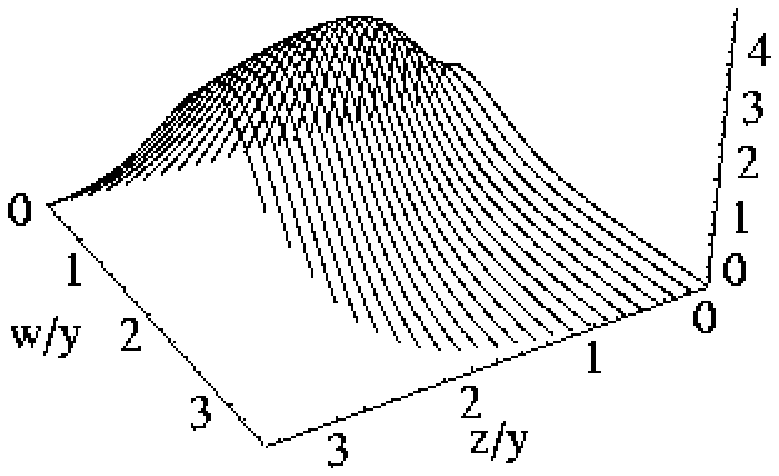}
\includegraphics[width=7.6cm,angle=0]{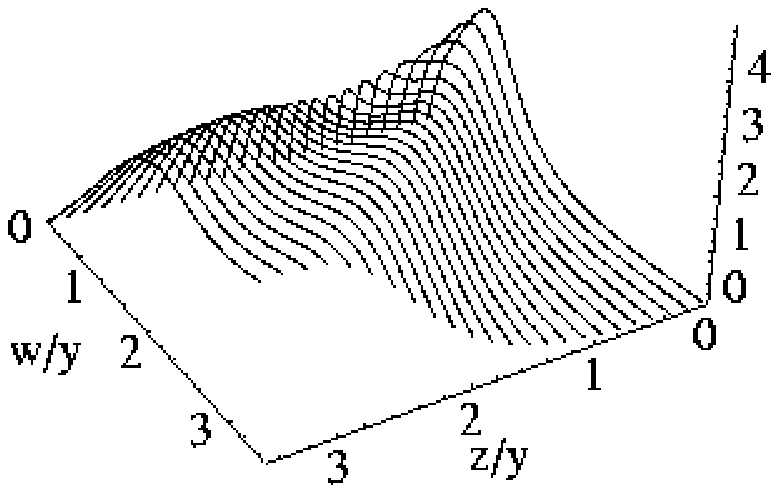}
\caption{The structure factors $S^{(e)} (z,w)/\nu$ (upper panel) and $S^{(o)} (z,w)/\nu$ (lower panel) at $y=\Gamma=10^{-3}$.}
\label{strucd}
%\end{center}
\end{figure}

Thus, we have shown that the propagating spin-density modes may occur in the non-diffusive regime, provided the SO coupling parameter exceeds the disorder broadening. In order to demonstrate that the latter condition is important we also plot in the Fig.~\ref{strucd} the both structure factors for the values $y=\Gamma=10^{-3}$. One can observe that now the peaks in the landscapes of $S^{(e,o)}$ are no longer sharply pronounced and well-resolved as compared to those shown in the Fig.~\ref{struc}. The trend  for decreasing ratio $y/\Gamma$ now becomes quite transparent: the resolution between the dispersions of the charge- and the spin-density modes is disappearing, and it completely vanishes in the dirty limit $y / \Gamma \ll 1$.

\section{Spin-density response to inhomogeneous longitudinal electric field}

It is natural to expect that the landscapes of the  components $\chi^{00}$, $\chi^{yy}$, and $\chi^{y0}$ would manifest qualitatively the same features as $S^{(e)}$, while the landscapes of  $\chi^{xx}$, $\chi^{zz}$, and $\chi^{zx}$ would have much in common with $S^{(o)}$. 

In this Section we briefly discuss the finite-$q$ behavior of the off-diagonal spin-charge component  $\chi^{y0}$. It is linked with  $s^{y}$-response $M_x^y$ to inhomogeneous longitudinal electric field $E_{q \omega} = - i q V^0_{q \omega}$ applied in the $x$-direction:
\be
M_x^y = \frac{i e\chi^{y0}}{2 q} = - \frac{e \nu}{4 k_F} \frac{2 w I^{y0}}{\Gamma z [(1-I^{00}) (1-I^{yy})-(I^{y0})^2]}.
\label{inmxy}
\ee

In the homogeneous limit we obtain the expression\cite{schwab,shnir1}
\be
\lim_{q \to 0} M_x^y = \frac{i e \nu}{2 k_F} \frac{y^3}{2 w (\tilde{w}^2-y^2)-i \Gamma y^2}.
\label{mxy}
\ee
In the clean regime $y \gg \Gamma$ Eq.~\eq{mxy} exhibits the pronounced resonance peak at $w \approx y$. Taking the stationary  limit $w \to 0$, one also recovers from \eq{mxy} the result of Ref.~\onlinecite{edel}.

\begin{figure}[t]
%\begin{center}
\includegraphics[width=7.6cm,angle=0]{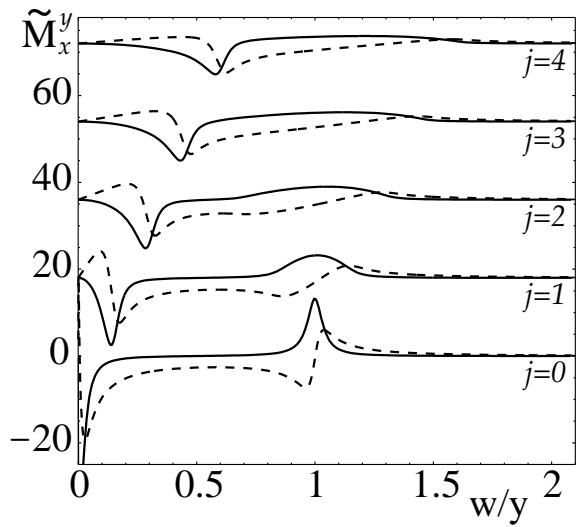}
\caption{The real (solid) and the imaginary (dashed) parts of $\widetilde{M}_x^y = (4 k_F/e \nu) M_x^y ( y\times \{0.01+0.15 j \},w)+18 j$ $[j=0,\ldots 4]$ at $y=10^{-3}$ and $\Gamma=0.5 \times 10^{-5}$.}
\label{mxypl}
%\end{center}
\end{figure}

The inhomogeneous response function \eq{inmxy} is expected to possess interesting features in the clean limit as well. In the Fig.~\ref{mxypl} we plot the normalized $M_x^y$ as a function of the frequency at different values of the momentum transfer. Solid and dashed curves correspond to the real and the imaginary parts of $M_x^y$, respectively. One can observe that with increasing $z$ the homogeneous peak at $w=y$ is getting suppressed, while there develops the other peak which has  the opposite sign and  is centered at $w=z$.

\section{Clean vs super-clean regimes}

Let us now present  {\it a posteriori} justification for distinguishing between the clean $y^2 < \Gamma < y$ and the super-clean $\Gamma < y^2$ regimes.  

As one can see from the expressions \eq{ice} and \eq{ico}, in the absence of the disorder broadening ($\Gamma \to 0$) the response functions $\chi$ would diverge at $w^{qc} =z$ and $w^{qc}_{\pm} =y \pm z$, i.e. at the same positions where the dispersions of the propagating modes in the clean regime have been observed in Sec.~\ref{propag}. In the limit $\Gamma \to 0$ these lines can be associated with the intrasubband and the intersubband boundaries of the Landau particle-hole excitation region which is defined by the condition ${\rm Im} \chi^{\alpha \alpha} \neq 0$. The fact that the boundaries in question have the linear form results from the quasiclassical approximation.

On the other hand, in Ref.~\onlinecite{plet1} it has been established that the intersubband Landau  region in the presence of the Rashba SO coupling is bounded by the parabolas $w_{\pm}=(y\pm z) \pm (y \pm z)^2$, and the imaginary part of the polarization operator does not diverge near them, but rather smoothly decays to zero. We note that the approach of  Ref.~\onlinecite{plet1} did not involve any sort of quasiclassical approximation, and was based on the exact evaluation of $\chi^{00}$ defined in Eq.~\eq{superclean}.

Thus, one can conclude that the quasiclassical approximation in the super-clean regime breaks down not only at $z, w \sim 1$, but also near the boundaries of the intersubband Landau region. In fact, one does need some amount of disorder broadening $\Gamma$ to regularize $\chi$ near the lines $w_{\pm}^{qc}= y \pm z$, and to thus justify the validity of the quasiclassical approximation in their neighborhood.  The quantitative criterion for this value of $\Gamma$ is
\be
\Gamma \geq \max |\Delta w_{\pm}|,
\ee
where $\Delta w_{\pm} = w_{\pm} -  w_{\pm}^{qc} \equiv \pm (y\pm z)^2$ is the mismatch between the actual and the quasiclassical Landau edges.

Since we are mostly concerned with the situations $z \sim y$, we obtain the condition $\Gamma > y^2$ which defines the clean regime. In the super-clean regime $\Gamma < y^2$ the quasiclassical approximation is no longer applicable, and therefore the study of the response functions \eq{superclean} requires a separate consideration similar to that of Ref.~\onlinecite{plet1}.

\section{Summary}

We have presented the compact analytical expressions for the density matrix response function of the 2DEG with Rashba SO coupling calculated at arbitrary ratio of $v_F q$, $\omega$, $\alpha_R k_F$, and $\tau^{-1}$. They describe the linear response of charge and spin densities to the spatially and temporally modulated fields in any regime of interest, namely: clean and dirty, diffusive and non-diffusive. The calculated expressions \eq{ice} and \eq{ico}, which define the density-density response to the nonuniform fields, as well determine the vertex corrections to the density-current and current-current response functions, and should prove useful in  evaluation of the latter in the nonuniform case.  

The structure factors containing the information about the quasiparticle modes have been considered. We have particularly discussed the dispersions and lifetimes of the propagating charge and spin-polarization modes occurring in the clean, non-diffusive regime, which is experimentally achievable by now. The spin-density response to the inhomogeneous longitudinal electric field has been studied in this regime as well.

The useful discussions with Matthias Eschrig, Vladimir Gritsev, and Alexander Shnirman are gratefully acknowledged. The work was supported by  the Deutsche For\-schungs\-ge\-mein\-schaft (DFG).

\appendix

\section{Dyson equation}
\label{dyssect}

A very general approach to the study of the quasi-particle dynamics is based on the Dyson equation\cite{rammer}
\be
(G_0^{-1} - \Sigma ) * G =1 ,
\label{dys}
\ee
where $G_0^{-1} = i \pd_t - H ({\bf k}) - V ({\bf r},t)$, and $V ({\bf r},t) = \sum_{\beta} \sigma^{\beta} V^{\beta}  ({\bf r}, t)$ is a time-dependent, spatially nonuniform and spin-dependent external potential. The Green's function $G$ and the self-energy $\Sigma$ are the matrices in the Keldysh space 
\be
G = \left( \begin{array}{cc} G_R & G_K \\ 0 & G_A\end{array}\right), \quad \Sigma = \left( \begin{array}{cc} G_R & G_K \\ 0 & G_A\end{array}\right),
\ee
which include the retarded (R), advanced (K), and the Keldysh (K) components. Eq. \eq{dys} can be equivalently rewritten as the set of equations
\beq
(G_0^{-1} - \Sigma_{R,A}) * G_{R,A} &=& 1, \label{graeq} \\
 (G_0^{-1}- \Sigma_{R}) * G_K -\Sigma_K * G_A &=& 0. \label{gkeq}
\eeq
They should be also complemented by the certain choice of the self-energy $\Sigma$. In the self-consistent Born approximation the latter is determined by the self-consistency  relation 
\be
\Sigma ({\bf r}_1 , t_1 ; {\bf r}'_1 , t'_1) = \frac{1}{m \tau } G ({\bf r}_1 , t_1 ; {\bf r}_1 , t'_1) \delta ({\bf r}_1 - {\bf r}'_1).
\label{sk}
\ee

There are different ways to treat the set of equations \eq{graeq}-\eq{sk}. One possibility is to apply the gradient expansion, and to thus derive the differential equation for the quasiclassical Keldysh function $G_K$\cite{shyt1}. Since $G_K$ is related to the distribution function, the solution of this equation determines the evolution of the charge and the spin densities\cite{schwab}.

Another possibility is to treat the equations \eq{graeq}-\eq{sk} perturbatively in $V$, and to derive the charge and the spin response functions. This approach is equivalent to the Kubo linear response theory, and in general it does not require the quasiclassical condition to be fulfilled. 

Let us expand $G$ and $\Sigma$ in a formal series of $V$:
\be
G \approx G^{(0)} + G^{(1)}, \quad \Sigma \approx \Sigma^{(0)} + \Sigma^{(1)}.
\ee   
One can find that the equilibrium functions $G^{(0)}$ and $\Sigma^{(0)}$ read
\beq
G^{(0)}_{R,A} &=& \left( E - H ({\bf k}) \pm \frac{i}{2 \tau} \right)^{-1},\\
\Sigma_{R,A}^{(0)} &=& \mp \frac{i}{2 \tau}, \label{self} \\
G^{(0)}_{K} &=& h (E) (G^{(0)}_{R} - G^{(0)}_{A}), \\
\Sigma^{(0)}_{K} &=& h (E) (\Sigma^{(0)}_{R} - \Sigma^{(0)}_{A}),
\eeq
where $h (E) = 1- 2 f (E) = \tanh \frac{E-E_F}{2 T}$, $f(E)$ is the Fermi distribution function, and $T$ is the temperature.

The deviation from the equilibrium is described by
\beq
G_K^{(1)} &=& G^{(0)}_{R} * V * G^{(0)}_{K} + G^{(0)}_{K} * V * G^{(0)}_{A} \nonumber \\
&+& G^{(0)}_{R} * \Sigma_K^{(1)} * G^{(0)}_{A}.
\label{gk1}
\eeq
The last term in this expression accounts the vertex corrections which are the same as  obtained by  the summation of the ladder diagrams in the diagrammatic approach.

Let us now introduce $g_K^{\alpha} = \frac12 {\rm Tr} \left( \sigma^{\alpha}G_K^{(1)}  \right)$ and $\sigma_K^{\alpha} = \frac12 {\rm Tr} \left( \sigma^{\alpha}\Sigma_K^{(1)}  \right)$, and rewrite \eq{gk1} in the Fourier representation. We obtain
\be
g_K^{\alpha} = g_K^{\alpha,I} +  g_K^{\alpha,II} + g_K^{\alpha,{\rm vertex}},
\label{escex}
\ee
where
\beq
& & g_K^{\alpha,I} = \sum_{\beta}  \left[ \textstyle{h (E + \frac{\omega}{2})} g^{\alpha \beta}_{AR} - \textstyle{h (E  - \frac{\omega}{2}}) g^{\alpha \beta}_{RA} \right] V^{\beta}_{q \omega}, \nonumber \\
& & g_K^{\alpha,II} = \sum_{\beta}  \left[ \textstyle{h (E - \frac{\omega}{2})} g^{\alpha \beta}_{RR} - \textstyle{h (E + \frac{\omega}{2}}) g^{\alpha \beta}_{AA} \right] V^{\beta}_{q \omega}, \nonumber \\
& & g_K^{\alpha,{\rm vertex}} = \sum_{\beta}  g^{\alpha \beta}_{RA} \sigma_{K}^{\beta}. \nonumber 
\eeq
The terms which appear in the above relations read
\beq
& & g_{R(A),R(A)}^{\alpha \beta} = \frac12 \sum_{\mu,\mu'} \textstyle{{\rm Tr} [\sigma^{\alpha} P_{\mu'} ({\bf k} +\frac{{\bf q}}{2})\sigma^{\beta} P_{\mu} ({\bf k} -\frac{{\bf q}}{2})]} \nonumber \\
& & \times \frac{1}{(E +\frac{\omega}{2} -\varepsilon^{\mu'}_{{\bf k}+\frac{{\bf q}}{2}} \pm \frac{i}{2 \tau}) (E -\frac{\omega}{2} -\varepsilon^{\mu}_{{\bf k}-\frac{{\bf q}}{2}} \pm\frac{i}{2 \tau})}.
\label{grr}
\eeq
The projective representation of \eq{grr} is due to the spectral decomposition of  $G^{(0)}_{R,A}$ similar to \eq{specdec}.

In order to find $\sigma^{\alpha}_K$ let us now integrate $g_K^{\alpha}$ over ${\bf k}$. Introducing the definitions
\be
I^{\alpha \beta}_{R (A),R (A)} = \frac{1}{m \tau} \int \frac{d^2 k}{(2 \pi)^2}
g_{R(A),R(A)}^{\alpha \beta} ,
\ee
and taking into account the self-consistency relation \eq{sk}, we obtain
\beq
\sigma_K &=& h (E+\omega/2) [I_{AR} - I_{RA}] V \nonumber \\
&+&  h (E-\omega/2) I_{RR} V - h (E+\omega/2) I_{AA} V \label{sigmak} \\
&+& [h (E+\omega/2) - h (E-\omega/2)] I_{RA} V + I_{RA} \sigma_K , \nonumber
\eeq
where the upper (spin and charge) indices are omitted for brevity. From this expression we can find $\sigma_K$, which is proportional to $V$, and consequently the linear response $\chi$ of the system:
\be
\rho^{\alpha}_{q\omega} = - m \tau \int \frac{d E}{2 \pi i} \sigma_K^{\alpha} = \sum_{\beta} \chi_{q\omega}^{\alpha \beta} V^{\beta}_{q\omega}.
\ee
One can show that the term in the first line of \eq{sigmak} is negligible, the second-line terms approximately give the constant term $\chi^{II}$, and the third line determines $\chi^{I}$ after the identifications $\varepsilon = E-E_F$ and $I_{RA} \equiv I$. Thus, at zero temperature the expression \eq{deps} is recovered. 

In the super-clean limit $\tau \to \infty$ the self-energy and the vertex corrections disappear, and we obtain from \eq{escex} the definition of the response functions
\beq
 \chi_{q\omega}^{\alpha \beta} &=& \sum_{\mu, \mu'} \int \frac{d^2 k}{(2 \pi)^2} \textstyle{{\rm Tr} [ \sigma^{\beta} P_{\mu} ({\bf k}) \sigma^{\alpha}  P_{\mu'} ({\bf k} +{\bf q})]}\nonumber \\
& \times & \frac{f (\varepsilon_{{\bf k}}^{\mu}) - f (\varepsilon_{{\bf k}+{\bf q}}^{\mu'})}{\omega +i 0 + \varepsilon_{{\bf k}}^{\mu} - \varepsilon_{{\bf k}+{\bf q}}^{\mu'}}.
\label{superclean}
\eeq

\section{Details of the  angle integration}
\label{integrat}

Let us explicitly list the components of ${\cal F}^{\alpha \beta}_{\mu \mu'} ({\bf k}, {\bf k}+ {\bf q})$ introduced in \eq{momint}. Along with the diagonal terms
\beq
{\cal F}^{00 (zz)}_{\mu \mu'} &=& 
\frac12 [1 \pm \cos (\phi_{{\bf k}+{\bf q}} - \phi_{{\bf k}})], \\
{\cal F}^{yy (xx)}_{\mu \mu'} &=& 
\frac12 [1 \pm \cos (\phi_{{\bf k}+{\bf q}} + \phi_{{\bf k}})],
\eeq
we have the off-diagonal spin-charge
\beq
{\cal F}^{0x}_{\mu \mu'} &=& {\cal F}^{x0}_{\mu \mu'} =  \frac12 (\mu \sin \phi_{{\bf k}} +\mu' \sin \phi_{{\bf k}+{\bf q}}), 
\\
{\cal F}^{0y}_{\mu \mu'} &=& {\cal F}^{y0}_{\mu \mu'} = - \frac12 (\mu \cos \phi_{{\bf k}} +\mu' \cos \phi_{{\bf k}+{\bf q}}), 
\\
{\cal F}^{0z}_{\mu \mu'} &=& - {\cal F}^{z0}_{\mu \mu'} = \frac{i}{2} \mu \mu' \sin (\phi_{{\bf k}+{\bf q}} - \phi_{{\bf k}}), 
\eeq
and spin-spin terms
\beq
{\cal F}^{yz}_{\mu \mu'} &=& - {\cal F}^{zy}_{\mu \mu'} =  \frac{i}{2} (\mu \sin \phi_{{\bf k}} - \mu' \sin \phi_{{\bf k}+{\bf q}}), 
\\
{\cal F}^{zx}_{\mu \mu'} &=& - {\cal F}^{xz}_{\mu \mu'} = -\frac{i}{2} (\mu \cos \phi_{{\bf k}} - \mu' \cos \phi_{{\bf k}+{\bf q}}), 
\\
{\cal F}^{xy}_{\mu \mu'} &=&  {\cal F}^{yx}_{\mu \mu'} = - \frac12 \mu \mu' \sin (\phi_{{\bf k}+{\bf q}} + \phi_{{\bf k}}).
\eeq

One can further elaborate these expressions using the relations
\beq
\cos (\phi_{{\bf k}+{\bf q}} - \phi_{{\bf q}}) &=& \frac{q+ k \cos \phi}{|{\bf k}+{\bf q}|}, \\ \sin (\phi_{{\bf k}+{\bf q}}-\phi_{{\bf q}}) &=& \frac{k \sin \phi}{|{\bf k}+{\bf q}|},
\eeq
where $\phi = \phi_{{\bf k}} - \phi_{{\bf q}}$.

Without the  loss of generality, we can choose the basis defined by ${\bf q} || {\bf e}_x$, or $\phi_{{\bf q}}=0$. This considerably simplifies our calculation due to the isotropy of the spectrum $\varepsilon_{{\bf k}}^{\mu}$ in the momentum space. In particular, one can observe that in this  basis the components $I^{0x}$, $I^{0z}$, $I^{yz}$, and $I^{xy}$ would identically vanish, since the corresponding integrals in \eq{angint} have the form
\be
\int_0^{2 \pi} d \phi \,\, \sin \phi \,\, F (\cos \phi) \equiv 0.
\ee

Let us make the following observation which allows us to perform an exact angle integration in \eq{angint} for the remnant components. Changing the dummy index $\mu' \to \mu \chi$, where $\chi =\pm$, and replacing $\sum_{\mu'} \to \sum_{\chi}$, we find after the latter summation that odd powers of $|{\bf k} + {\bf q}|$ disappear, and the integrand becomes a rational function of $\cos \phi$ (cf. Ref.~\onlinecite{plet1}). It only remains to expand it into simple fractions, and to apply the standard 
integral
\be
\frac{1}{2 \pi} \int_0^{2 \pi} \frac{d \phi}{1+b \cos \phi} = \frac{1}{\sqrt{1-b^2}},
\ee
where $b$ is the arbitrary complex number. Thus, after tedious, but straightforward, calculation one can recover Eqs. \eq{ice} and \eq{ico}.

We also note that the expression for the components of $I$ in the arbitrary basis can be obtained by the orthogonal transformation $\tilde{I}  = O_{{\bf q}} I  O_{{\bf q}}^{-1}$, where 
\be
 O_{{\bf q}} = \left( \begin{array}{cccc} 
1 & 0 & 0 & 0 \\
0 & \cos \tilde{\phi}_{{\bf q}}& -\sin \tilde{\phi}_{{\bf q}}& 0 \\
0 & \sin \tilde{\phi}_{{\bf q}}& \cos \tilde{\phi}_{{\bf q}}& 0 \\
0 & 0 & 0 & 1 
\end{array} \right),
\ee
and $\tilde{\phi}_{{\bf q}}$ is the angle of the vector ${\bf q}$ in the new basis.

\end{document}